\documentclass[useAMS,usenatbib]{mn2e}
\usepackage{graphicx,subfigure} % subfigure is needed to put multiple plots in one figure
\usepackage{amsmath}

\newcommand\lsim{\la}
\newcommand\gsim{\ga}

\newcommand \mum{ \mu{\rm m} }
\newcommand \amin{ a_{\rm min} }
\newcommand \amax{ a_{\rm max} }
\newcommand \charsig{ \overset{\sim}{\sigma} }

\newcommand \NH{N_{\rm H}}
\newcommand \tauext {\tau_{\rm ext}}
\newcommand \tausca {\tau_{\rm sca}}
\newcommand \zals {z_a}

\newcommand \gcmcu{g~cm$^{-3}$}
\newcommand \column{cm$^{-2}$}

\newcommand \tbabs{\texttt{tbabs}}
\newcommand \phabs{\texttt{phabs}}
\newcommand \tbnew{\texttt{TBnew}}

\newcommand \Chandra {{\sl Chandra}}
\newcommand \Swift {{\sl Swift}}
\newcommand \XMM {{\sl XMM-Newton}}
\newcommand \Suzaku {{\sl Suzaku}}
\newcommand \nustar {{\sl NuSTAR}}

\begin{document}

\title[The dust scattering component of X-ray extinction]{The dust scattering component of X-ray extinction: \\ Effects on continuum fitting and high-resolution absorption edge structure}
\author[Corrales et. al.]
{
L.~R.~Corrales$^1$, J.~Garc\'ia$^2$, J.~Wilms$^3$, F.~Baganoff$^1$ \\
$^1$MIT Kavli Institute for Astrophysics and Space Research\\
$^2$Harvard Center for Astrophysics\\
$^3$University of Erlangen-Nuremberg
}

% Rejected from abstract because MNRAS as a 250 word limit
%For the most obscured sight lines, $\NH \gsim 10^{23}$~cm$^{-2}$, the photon index and unabsorbed source flux will be underestimated by 10-50\%.  
%
%For the case of dusty material enshrouding young stars or winds from X-ray binary stellar companions, only absorption is needed to model the intrinsic extinction.

\maketitle
\begin{abstract}
Small angle scattering by dust grains causes a significant contribution to the total interstellar extinction for any X-ray instrument with sub-arcminute resolution (\Chandra, \Swift, \XMM).  However, the dust scattering component is not included in the current absorption models: \phabs, \tbabs, and \tbnew.  We simulate a large number of \Chandra\ spectra to explore the bias in the spectral fit and $\NH$ measurements obtained without including extinction from dust scattering.  We find that without incorporating dust scattering, the measured $\NH$ will be too large by a baseline level of 25\%.  This effect is modulated by the imaging resolution of the telescope, because some amount of unresolved scattered light will be captured within the aperture used to extract point source information.  In high resolution spectroscopy, dust scattering significantly enhances the total extinction optical depth and the shape of the photoelectric absorption edges.  We focus in particular on the Fe-L edge at 0.7~keV, showing that the total extinction template fits well to the high resolution spectrum of three X-ray binaries from the \Chandra\ archive: GX 9+9, XTE J1817-330, and Cyg X-1.  In cases where dust is intrinsic to the source, a covering factor based on the angular extent of the dusty material must be applied to the extinction curve, regardless of angular imaging resolution.  This approach will be particularly relevant for dust in quasar absorption line systems and might constrain clump sizes in active galactic nuclei.
\end{abstract}


%%- - - - - - - - - - - - - - - - - - - - - - - - - - - - - - - - - - - - - - - - - - - - - - - - - - - %%
\section{Introduction}
\label{sec:Introduction}

% 1. Dust scattering contributes a comparable amount of extinction to dust absorption of X-ray light (Fig 1)
% 2. Majority of ISM metals are locked up in dust, with exception of Oxygen, Nitrogen, Neon (Fig 2)

\begin{figure*}
\begin{center}
	\includegraphics[width=\textwidth]{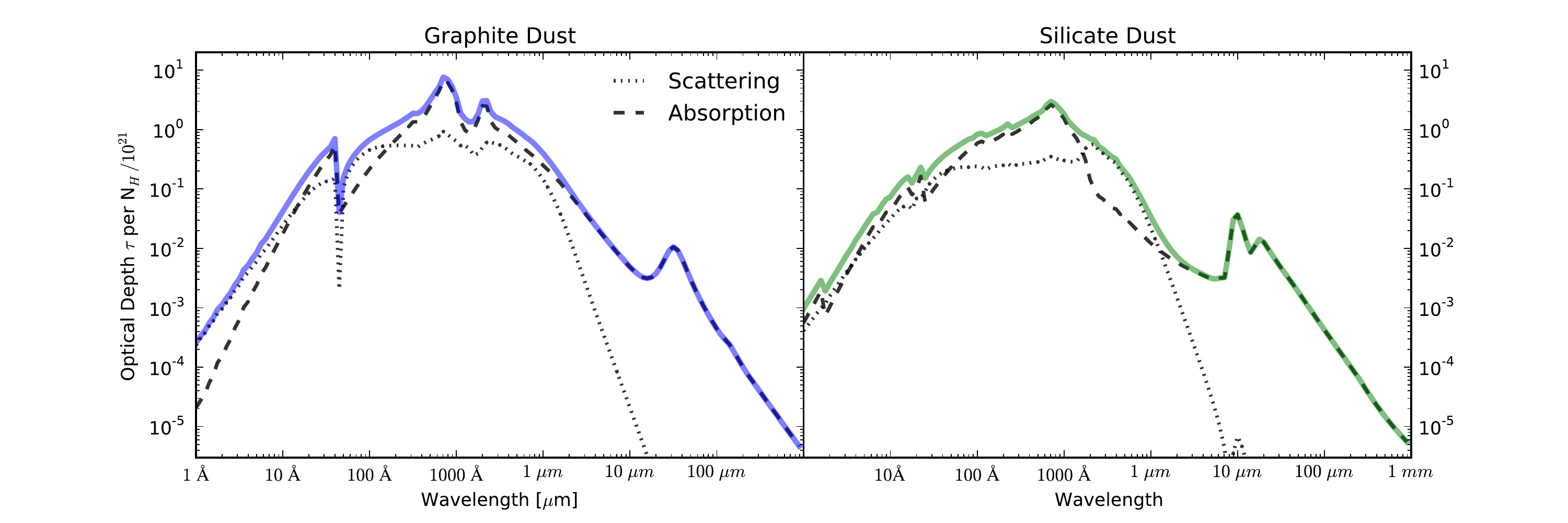}
	\caption{The total extinction opacity for a power law distribution of dust grain sizes ($dn/da \propto a^{-3.5}$ where the grain radius $a$ spans 0.005~$\mum$ to 0.25~$\mum$) using a dust-to-gas mass ratio typical of the Milky Way.  In each plot, it is assumed that all of the interstellar dust is locked up in either graphite grains (left) or silicate grains (right), using the optical constants  from \citet{Draine2003b}.}
	\label{fig:ISMextinction}
\end{center}
\end{figure*}

The relative transparency of the Milky Way interstellar medium (ISM) to X-rays allows us to examine the signatures of absorption and scattering over several orders of magnitude of ISM column density ($\NH \sim 10^{20} - 10^{23}$~cm$^{-2}$).  Figure~\ref{fig:ISMextinction} shows the relative contributions of scattering and absorption for a theoretical extinction curve spanning the IR to the X-ray, under the assumption of a power law distribution of dust grains of either graphite or silicate mineralogy.  Below 10~\AA\ ($E \geq 1$~keV), scattering from dust contributes equal parts or more to absorption from dust.  
However, the extinction models commonly applied by X-ray astronomers to fit the continuum -- \phabs, \tbabs, and \tbnew\ -- account only for ISM absorption via the photoelectric effect.  To be fully accurate, the spectral models for obscured X-ray point sources need to include the extinction component from dust scattering.

Dust scatters X-ray light through small angles on the order of $10.4' (a/0.1~\mum)^{-1} (E/{\rm keV})^{-1}$ \citep{Overbeck1965,Hayakawa1970,MG1986,SD1998}.  For detectors with high enough imaging resolution (\Chandra, \Swift, and \XMM) dust scattering halos are ubiquitously seen for objects with $\NH \gsim 10^{22}$~cm$^{-2}$ \citep{Rolf1983,PS1995,VS2015}.  X-ray imaging detectors with much lower angular resolution (e.g.,~\Suzaku, 
\nustar, and {\sl Astro-H}) will have a scattering halo that is enclosed by the typical point source spectral extraction aperture, which will recapture the scattering component and, in effect, mitigate the extinction contribution from dust \citep[e.g.,][]{Nowak2011}.
For instruments with sufficient resolution, we will evaluate the level of systematic bias that occurs when dust scattering is not included in the ISM extinction model used to determine gas column ($\NH$) in Section~\ref{sec:TBdust}.

The dust scattering cross-section also exhibits resonances that correspond to the photoelectric absorption edges.  This will alter the appearance of the edge, particularly on the soft-energy side.  We will show that an absorption plus scattering Fe-L edge template fits well to three Galactic binaries (Section~\ref{sec:HighResolution}).  
%
%To conclude, we will evaluate the extent to which the next generation of X-ray telescopes will be affected by dust scattering and to what extent they will be able to determine the composition of interstellar dust.  
To conclude, we will evaluate the extent to which dust scattering will affect sources with intrinsic obscuration by dust.  We examine this in the context of several science problems -- obscuration by clumpy materials in active galactic nuclei (AGN) and dust in quasar absorption line systems (Section~\ref{sec:Discussion}).

%%- - - - - - - - - - - - - - - - - - - - - - - - - - - - - - - - - - - - - - - - - - - - - - - - - - - %%
\section{Contribution of dust scattering to the total ISM extinction and continuum models}
\label{sec:TBdust}

We use the canonical power-law grain size distribution of \citet[][hereafter MRN]{MRN1977} as a preliminary model for interstellar dust.  MRN found that a mix of graphite and silicate grains can be used to reproduce the ISM extinction curves for the Milky Way.  We use optical constants for 0.1~$\mum$ sized graphite and silicate grains, covering wavelengths from the far-infrared to X-ray \citep[][and references therein]{Draine2003b}, and calculate the total extinction opacity by integrating the total Mie extinction efficiency over the grain size distribution:
\begin{equation}
	\label{eq:TauExt}
	\tauext(E) = \int_{\amin}^{\amax} Q_{\rm ext}^{\rm Mie}(a, E) \pi a^2 \frac{dn}{da} da
\end{equation}
where $dn/da \propto a^{-3.5}$, $\amin = 5$~nm, and $\amax = 0.3~\mum$ for both grain compositions.  The normalization for the dust grain size distribution depends purely on the total dust mass.  For Galactic objects, we estimate the dust mass by assuming a constant dust-to-gas mass ratio of 0.009 and a dust grain material density of 2.2~\gcmcu\ and 3.8~\gcmcu\ for graphite and silicate grains, respectively.  
We find that a 60/40 mix of silicate to graphite grains roughly approximates the $R_V=3.1$ extinction curve, which represents the average for diffuse Milky Way ISM.  This dust component mixture and grain size distribution constitute our fiducial dust abundance model for the Milky Way.  
We use the Mie scattering algorithms of \citet{BHmie}, implemented in python \citep{dustrepo}.

\begin{figure*}
\begin{center}
	\includegraphics[width=\textwidth]{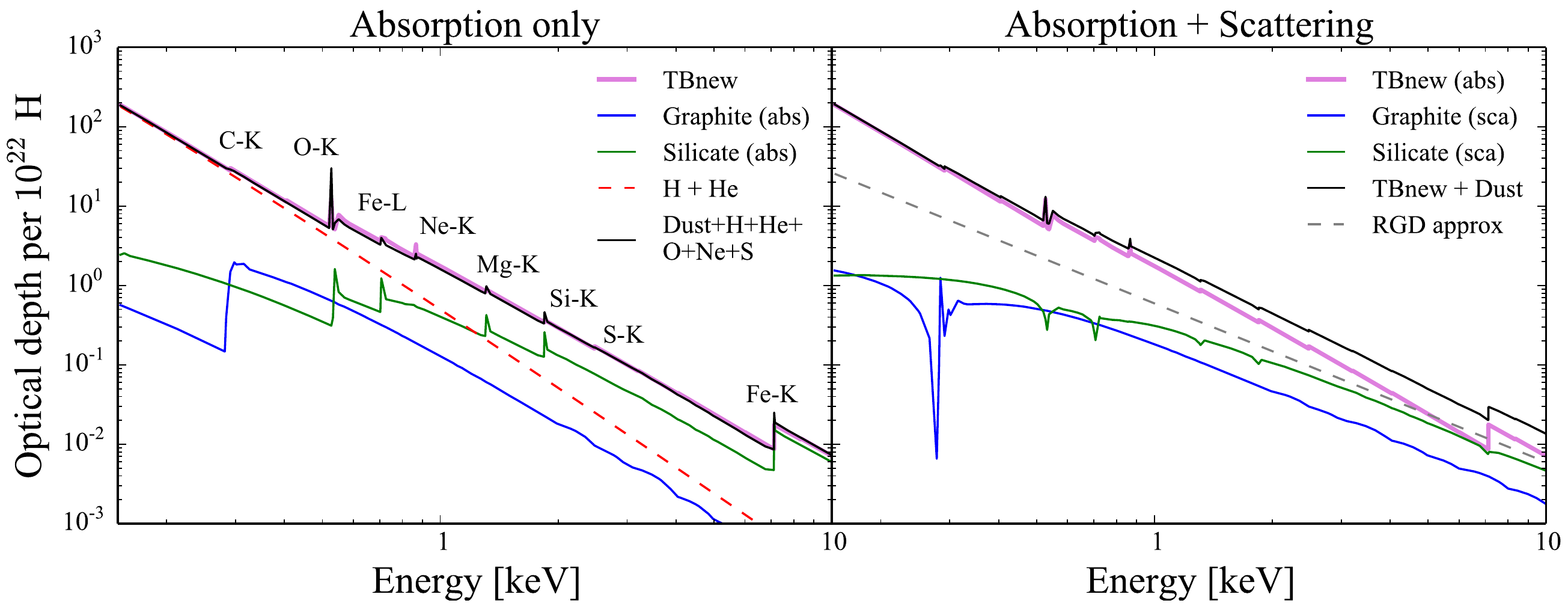}
	\caption{
	(Left) The contribution of a 60/40 silicate (green) and graphite (blue) dust to the \tbnew\ absorption opacity (magenta) for an ISM column of $10^{22}$~\column.
	(Right) The combined and absorption and scattering components of a 60/40 silicate to graphite mix give the total extinction opacity for an ISM column of $10^{22}$~\column (black).  The Rayleigh-Gans plus Drude approximation, used often in X-ray studies of dust scattering, is shown for reference (grey dashed line).
	}
	\label{fig:TBnew}
\end{center}
\end{figure*}

Figure~\ref{fig:TBnew} (left) shows the X-ray optical depth of the ISM for absorption only, using \tbnew\  \citep{Hanke2009}\footnote{http://pulsar.sternwarte.uni-erlangen.de/wilms/research/tbabs}.  Below 0.5~keV, photoelectric absorption from hydrogen and helium dominates.  Due to its relative abundance and absorption edge depth, the photoelectric absorption edge from oxygen causes a dramatic increase in the absorption cross-section above pure H and He.  A large fraction of interstellar metals are locked up in dust grains, which are relatively transparent to X-rays.\footnote{Due to shielding, the total metal column available for photoelectric absorption is reduced for the largest dust grains, but this effect is most noticeable for ISM regions containing a significant abundance of large grains, e.g., singularly sized $0.3~\mum$ grains \citep{Wilms2000}.}  Thus almost all the absorption from Si, Mg, and Fe can be accounted for by dust.

However, several abundant elements with strong absorption edges do not easily condense into dust grains.  A large fraction of C, O, and S remain in gas form.  Since C absorption is not as strong as H and He, we only needed to include absorption from gaseous O and S to reach parity with the \tbnew\ absorption model.  We adopt the elemental depletion factors of \citet{Wilms2000} for oxygen, neon, and sulfur -- for which 60\%, 100\%, and 60\% of the respective neutral atoms are expected to be in the gas phase.  The absorptive cross section for dust in Figure~\ref{fig:TBnew} (left) uses the fiducial 60/40 mix of silicate to graphite dust.  With the remaining abundant elements added, we find a relatively good agreement between the chosen dust mix and \tbnew.  One thing to note is that this implies that all neutral interstellar iron is locked up in dust grains, contrary to the 30\% value of gaseous iron used by \citet{Wilms2000}.  This result is consistent with the neutral Fe-L edge measurements from the foreground ISM of X-ray binaries, which match the cross-sections for solid iron \citep{Juett2006, Gatuzz2015}.  Meanwhile, there are no reliable experimental or theoretical cross sections for atomic Fe in the literature, so its contribution to the Fe-L shell region has not been properly studied.

\begin{figure}
\begin{center}
	\includegraphics[width=\columnwidth]{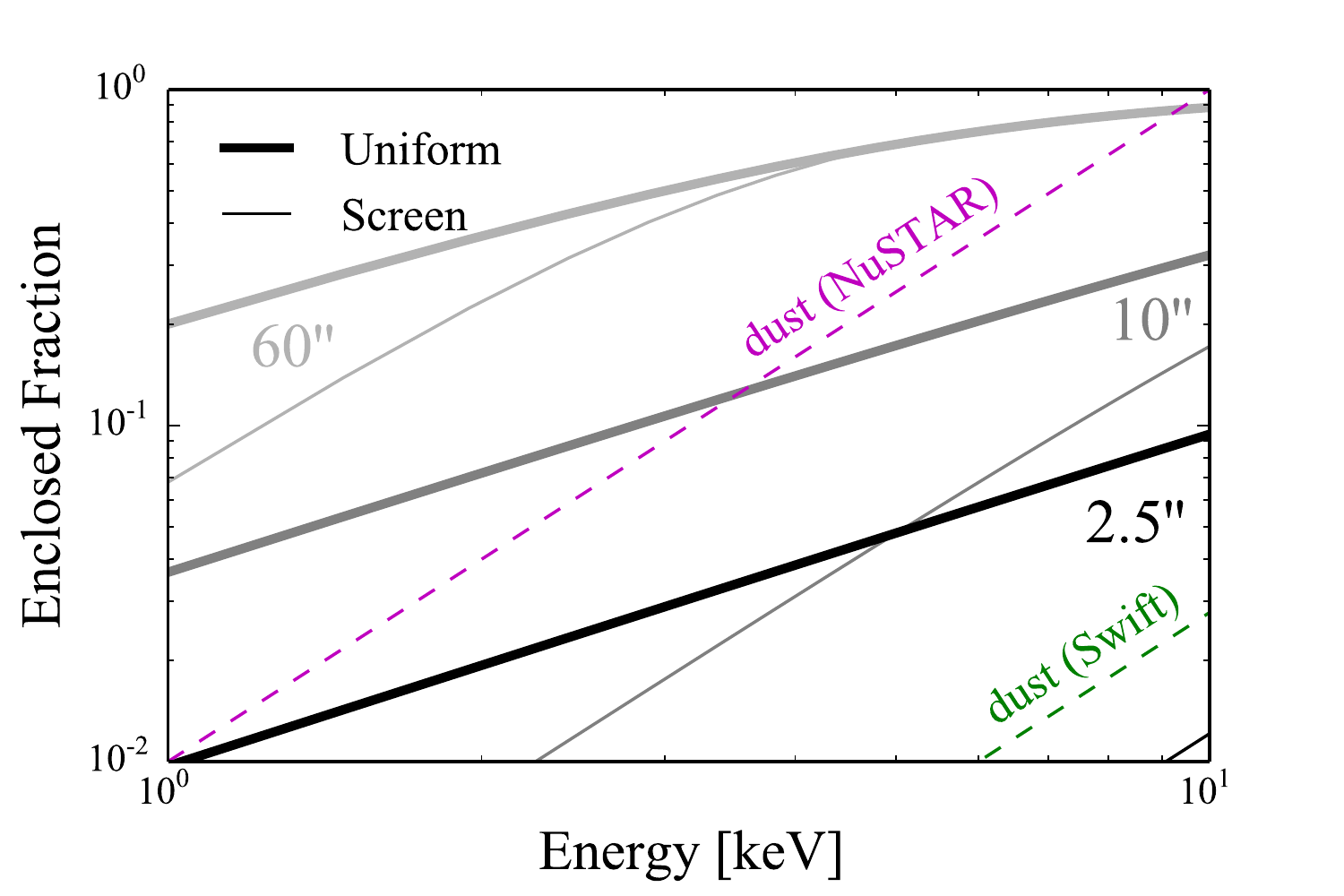}
	\caption{
	Fraction of the dust scattering halo enclosed by a source extraction region of various radii.  The thick lines show the enclosed fraction from a scattering halo calculated from dust distributed uniformly along the line of sight.  The thin lines show the same but for a situation where all the dust located in a wall exactly half way between the X-ray source and observer.  The $60''$ curves are indicative of low resolution telescopes such as \nustar\ and {\sl Astro-H}, the $10''$ curves are indicative for \XMM\ and \Swift, and the $2.5''$ curve is relevant for the default \Chandra\ HETG source extraction width.  
	For comparison, the XSPEC model \texttt{dust} assumes that all of the scattered flux is spread uniformly in a disk with radius $\propto E^{-1}$, which implies an $E^2$ dependence for the enclosed fraction.  The dashed lines show what such a geometry would predict for \nustar\ and \Swift.
	}
	\label{fig:EnclosedFraction}
\end{center}
\end{figure}

Using our fiducial dust abundance model, Figure~\ref{fig:TBnew} (right) shows the total extinction optical depth of the ISM by summing \tbnew\ absorption with the Mie scattering cross-sections of the dust grains.  This plot emphasizes how ISM extinction due to dust scattering causes a significant departure from the pure absorption model at moderate energies, $1-10$~keV.  However, due to the relatively poor imaging resolution of most X-ray telescopes, some fraction of the scattered light will be confused with non-scattered point source light.  The degree to which the scattered light is ``recaptured'' depends on the size of the region used to extract a point source spectrum {\sl and} the shape of the scattering halo surface brightness profile (Figure~\ref{fig:EnclosedFraction}).  It suffices to say that most scattering halos around Galactic X-ray binaries are dominated by a single dust wall \citep{VS2015}, and so \XMM\  and \Swift\ should lose $>$90\% of dust scattered light over their bandpass.  Around the Fe-K band, \nustar\ will lose 30\% of the dust scattered light, but this will only affect Fe-K features on the few percent level for the most highly obscured sight lines (e.g.,~the Galactic Center, which has $\NH \approx 10^{23}$~cm$^{-2}$).

%% Talk about fraction dust scattering halo recaptured by telescope

For reference, the Rayleigh-Gans plus Drude (RGD) approximation, used in many studies of X-ray scattering halos \citep[e.g.,][]{MG1986,SD1998,CorralesCygX3}, is plotted along side the more accurate Mie scattering cross-sections in Figure~\ref{fig:TBnew} (right).  The power of the RGD approximation is that it does not depend specifically on grain composition, only the average grain material density.  It follows an $E^{-2}$ energy dependence, with an optical depth $\tausca (1~{\rm keV}) \approx 0.05$ per $\NH = 10^{21}$~\column, calculated from our fiducial MRN distribution.  This value is consistent with the results of \citet{PS1995}, who used all the X-ray scattering halos available from the ROSAT all-sky survey to determine the average dust scattering parameters of the ISM. 
%
%Recently, \citet{Nowak2012} renormalized this value based on the different ISM abundances assumed between \texttt{wabs} (used by \citet{PS1995}) and \tbnew, which yields a 30\% difference in the scaling relations between $\tausca$ and $\NH$.  
%%
%Reconciling these differences is an open problem and will be visited in a later work.

\begin{figure*}
\begin{center}
	\includegraphics[width=\textwidth]{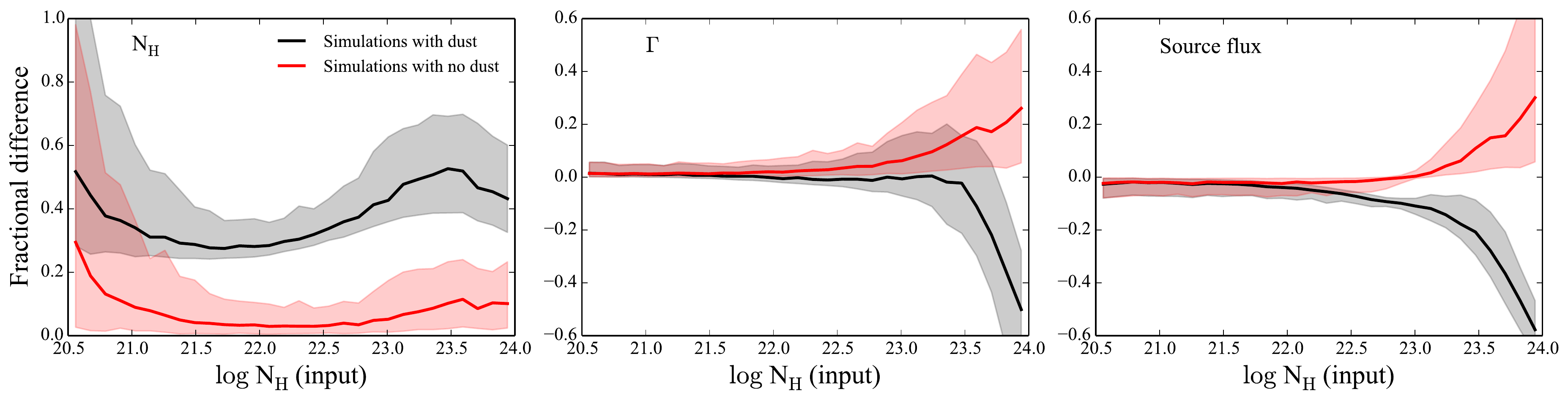}
	\caption{
	Fractional difference between fit and input parameters $(X_{\rm fit} - X_{\rm inp})/X_{\rm inp}$ as a function of input $\NH$ value.  (Left) Without dust extinction, fit $\NH$ values are systematically too high by a baseline level of 25\% and up to 50\% for input $\NH > 10^{23}$~\column.  (Middle) When absorption begins to affect the high energy end of the energy range covered, the photon index $\Gamma$ can be underestimated by a factor of 50\% when $\NH \gsim 3 \times 10^{23}$~\column.  (Right) The inaccuracy of the source parameters will lead the unattenuated source flux to be underestimated by 10-50\% for $\NH > 10^{23}$~\column.
	}
	\label{fig:Simulations}
\end{center}
\end{figure*}

To estimate the systematic effect of neglecting dust scattering from models of ISM extinction, we simulated a large number of \Chandra\ ACIS-I spectra using the Interactive Spectral Interpretation System \citep[ISIS,][]{ISISsoftware}.  We assume a power law emission spectrum for the source, under the influence of extinction from neutral ISM metals and dust scattering:  
\begin{equation}
\label{eq:Ftrue}
	F_{\rm ps}^{\rm inp} \propto E^{-\Gamma} \ \exp( -\tau_{\rm abs} - \tau_{\rm sca} )
\end{equation}
where $\tau_{\rm abs}$ comes from \tbnew\ and $\tau_{\rm sca} = 0.05 (\NH/10^{21} {\rm cm}^{-2})$.  
This equation assumes the worst-case scenario: 100\% loss of the dust scattered light from the source extraction region, which is valid for \Chandra\ observations with a dust wall at moderate distances between the source and observer (Figure~\ref{fig:EnclosedFraction}).  Then we fit the spectra without dust extinction: 
\begin{equation}
\label{eq:Ffit}
	F_{\rm ps}^{\rm fit} \propto E^{-\Gamma} \ \exp( -\tau_{\rm abs} )
\end{equation}
using standard $\chi^2$ statistics.  In order to gauge systematic effects that come from the fitting method and not the extinction model, we also simulate spectra according to Equation~\ref{eq:Ffit} and then fit them.  
The modeled spectra were chosen to have 2-10~keV apparent fluxes uniformly distributed in log space between 0.1 and 10~mCrab ($2.5 \times 10^{-12} - 2.5 \times 10^{-10}$~erg~cm$^{-2}$~s$^{-1}$).  In addition, photon indices were chosen uniformly between 1 and 3, and exposure times were chosen uniformly between 15 and 100~ks.
These parameters were chosen to be representative of typical \Chandra\ observations of bright Galactic X-ray binaries and some extragalactic point sources.

Figure~\ref{fig:Simulations} shows the systematic offsets that come from the fit method (red) versus the effect of not including the scattering in the ISM extinction model (black).  Interestingly, we find that neglecting dust scattering causes $\NH$ values to be systematically overestimated by a baseline factor of  25\% for all objects.  This bias increases to up to 50\% for the most obscured sight lines.  
In addition, the unattenuated source flux calculated from such models can be 10-50\% too low when $\NH >$~few~$\times 10^{23}$~\column.  More specific examples of how dust extinction can alter key science results in the study of X-ray binaries are given in Smith et al.~(2016, in prep).

%% Add to figure: dashed black line that shows continuum model

%%- - - - - - - - - - - - - - - - - - - - - - - - - - - - - - - - - - - - - - - - - - - - - - - - - - - %%
\section{The effects of dust scattering at high resolution}
\label{sec:HighResolution}

The dust scattering cross section can alter the appearance of photoelectric absorption edges for instruments with sufficient imaging and spectral resolution.  The resonances that lead to an absorption edge also cause resonances in the scattering cross section that will alter the shape of an absorption edge feature and, in particular, will add a lip to the low-energy side of the edge.  Judging from Figure~\ref{fig:TBnew} (right), these features will be most dramatic for the C-K (0.3~keV), O-K (0.5~keV), and Fe-L edges (0.7~keV).
Since a large fraction of interstellar C and O is likely in gas form (about 50\% and 60\%, respectively), full fits to the absorption edge region will need additional contributions from gas phase neutral and low-ionization states of those elements.  Where ionization might be relevant, we recommend the \texttt{ISMabs} model of \citet{Gatuzz2015}.  From here forward we focus in particular on interstellar Fe.

\begin{figure}
\begin{center}
	\includegraphics[width=\columnwidth]{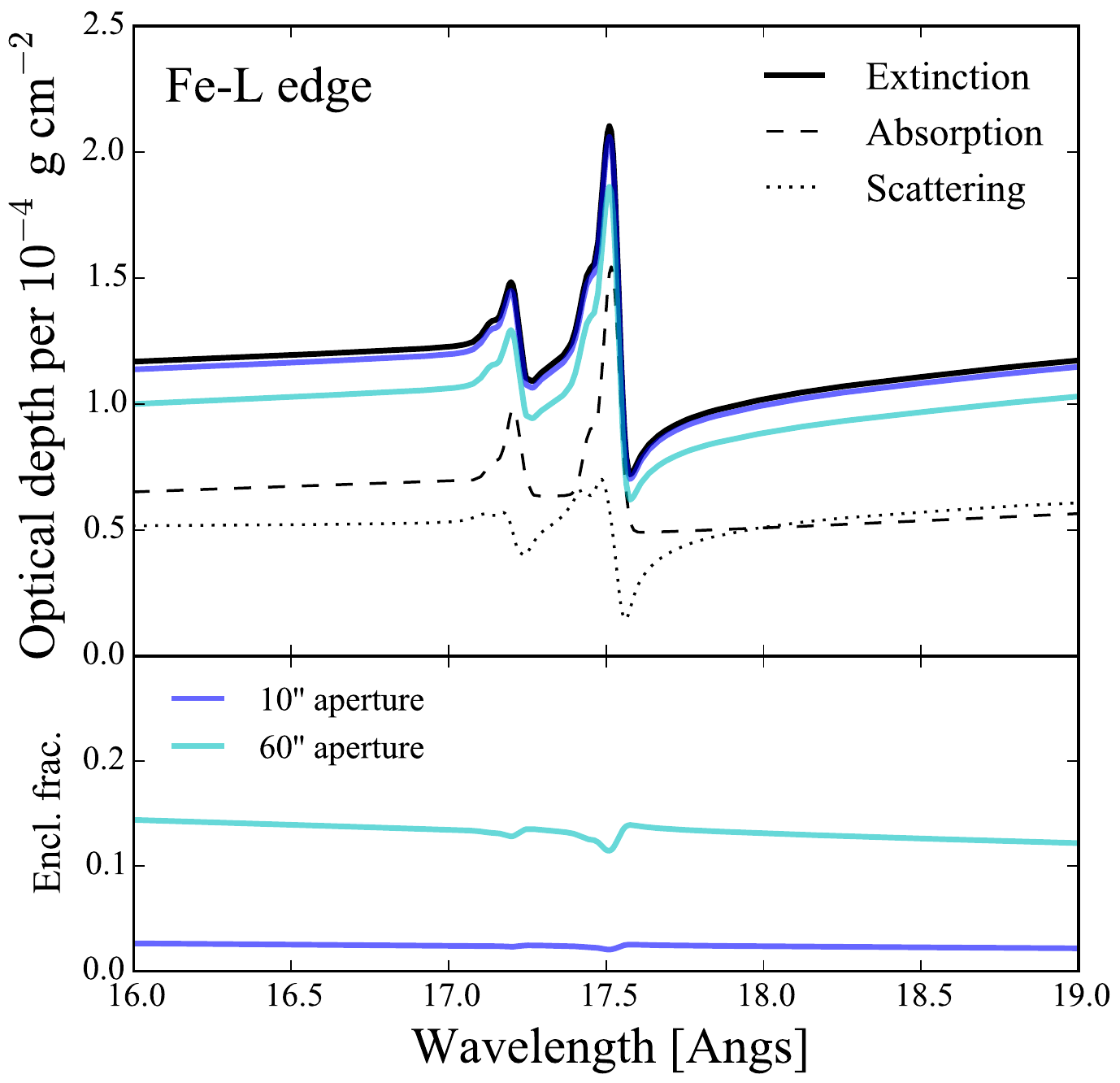}
	\caption{
	The optical depth of the Fe-L edge using optical constants from \citet{Draine2003b}, which are based on lab absorption measurements of olivine \citep{vanAken2002}.  Under our fiducial dust abundance assumptions, the dust mass column of $M_d = 10^{-4}$~g~cm$^{-2}$ is equivalent to $\NH \approx 10^{22}$~\column.  Because some of the scattered light will be captured within the aperture used to extract the source spectrum, the Fe-L edge extinction cross-section will depend on telescope resolution.  For a $10''$ ($60''$) radius aperture, the enclosed fraction of scattered light is about 2.5\% (15\%), assuming that the dust is uniformly distributed along the line of sight. 
	}
	\label{fig:FeL}
\end{center}
\end{figure}

Figure~\ref{fig:FeL} shows a close up of the Fe-L edge, computed with Mie scattering and our MRN model of silicate dust.  We use the optical constants of \citet{Draine2003b}, which are based on the lab measurements of olivine absorption by \citet{vanAken2002}.  We show here the cross-section for a given ferric dust mass column in units of $10^{-4}$~g~cm$^{-2}$.  Under our fiducial dust abundance assumptions, the corresponding hydrogen column is $\NH \approx 10^{22}$~cm$^{-2}$.  However, recent work by Schulz et al.~(2016, in prep) find significant departure from solar metallicity abundances for ISM silicates.  For this reason, we opt to describe the dust cross-section in terms of mass column, because the conversion from $\NH$ values depends on strong assumptions about ISM metallicity and depletion factors.  For instance, we find that using the ISM abundance and depletion values from Table~1 of \citet{Wilms2000} implies a dust-to-gas mass ratio of $6.5\times10^{-3}$; if all interstellar Fe is depleted into dust grains, the dust-to-gas mass ratio raises to $6.9\times10^{-3}$.  In addition, converting from dust mass to total Fe column requires knowledge of the ferric dust mineralogy. 
 %
%\citet{Gatuzz2015} were able to do this because the Fe-L cross section they used came from pure metallic iron \citep{Kortright2002}.  
If the Fe is in olivine dust, (Mg,Fe$^{2+}$)$_2$SiO$_4$, then the conversion to Fe column would depend directly on the Fe to Si abundance ratio and depletion factors.

\begin{figure}
\begin{center}
	\includegraphics[width=\columnwidth]{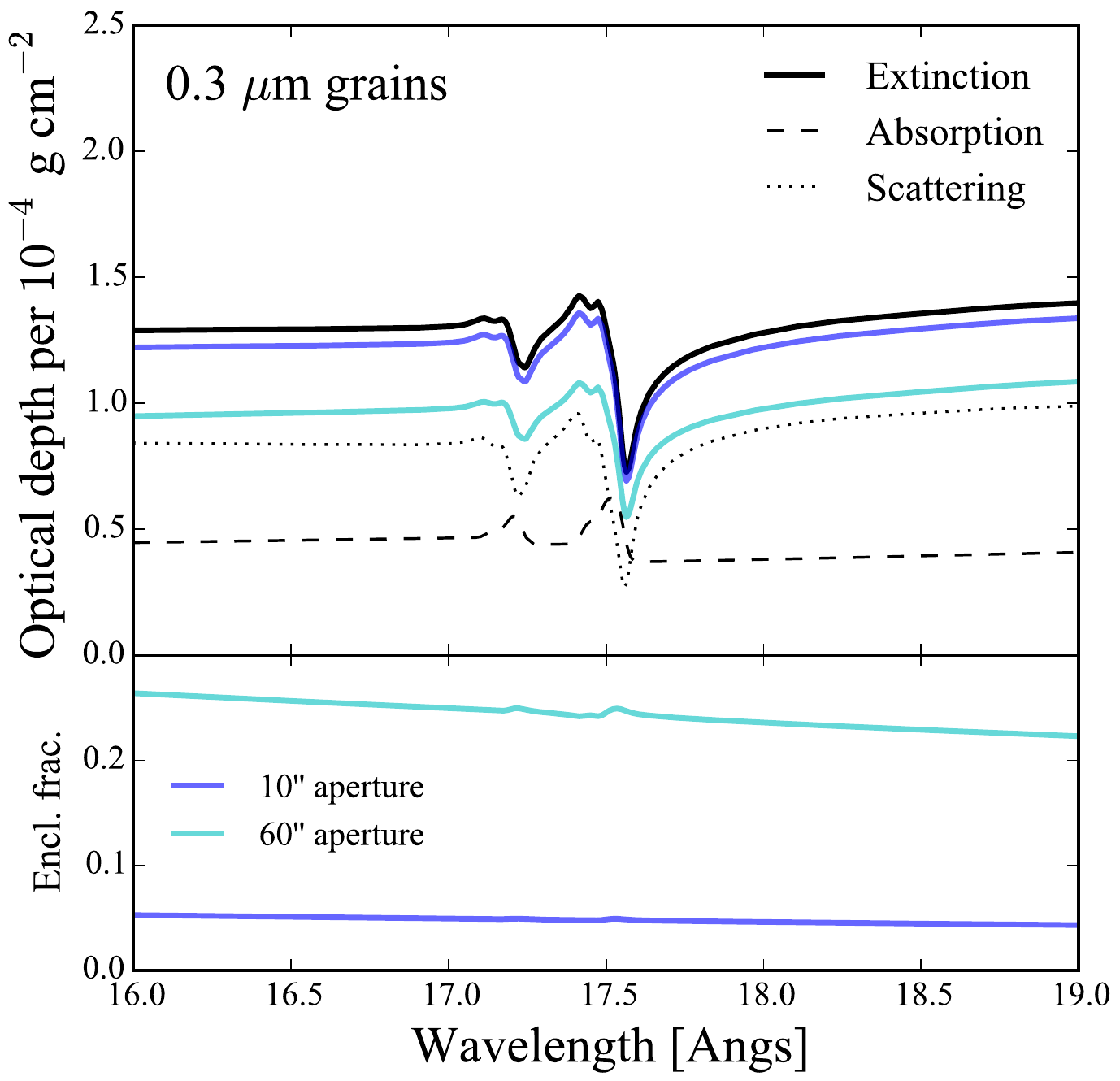}
	\caption{
	The optical depth of the Fe-L edge using the same models as Figure~\ref{fig:FeL}, except that all grains are $0.3~\mum$ in radius.  Because large dust grains produce more compact scattering halos, the enclosed fraction of scattered light for a $10''$ ($60''$) radius aperture is about 5\% (25\%).
	}
	\label{fig:FeL_big}
\end{center}
\end{figure}

The photoelectric absorption edges will also change according to the dust grain size distribution, grain shape \citep{HD2015}, grain mineralogy \citep{Lee2009,Lee2010}, and imaging resolution of the relevant telescope as described by Figure~\ref{fig:EnclosedFraction}.  The blue curves in Figure~\ref{fig:FeL} show how the extinction cross-section at the Fe-L edge might change for instruments with $10''$ resolution (e.g.,~\XMM\  or {\sl Swift}) and $60''$ resolution (e.g.,~\nustar\  or the upcoming {\sl Astro-H} mission).  If we go to the extreme of assuming that all interstellar dust grains are $0.3~\mum$ in size, Figure~\ref{fig:FeL_big} illustrates how the depth of the photoelectric absorption edge decreases dramatically.  This is due to shielding, as strong absorption prevents X-rays from penetrating the inner portions of the dust grain, and a smaller fraction of the total metal column contributes to the X-ray absorption edge.  The dust grain size distribution and mineralogy give a wide variety of parameters that may be tuned to match the observed absorption edge structure.  Ultimately, more power to discern between different grain models will come with the next generation of X-ray telescopes that have higher energy resolution and larger effective area ({\sl Astro-H}, {\sl Athena}, and the proposed X-ray Surveyor mission).

\begin{figure*}
\begin{center}
	\includegraphics[width=\textwidth]{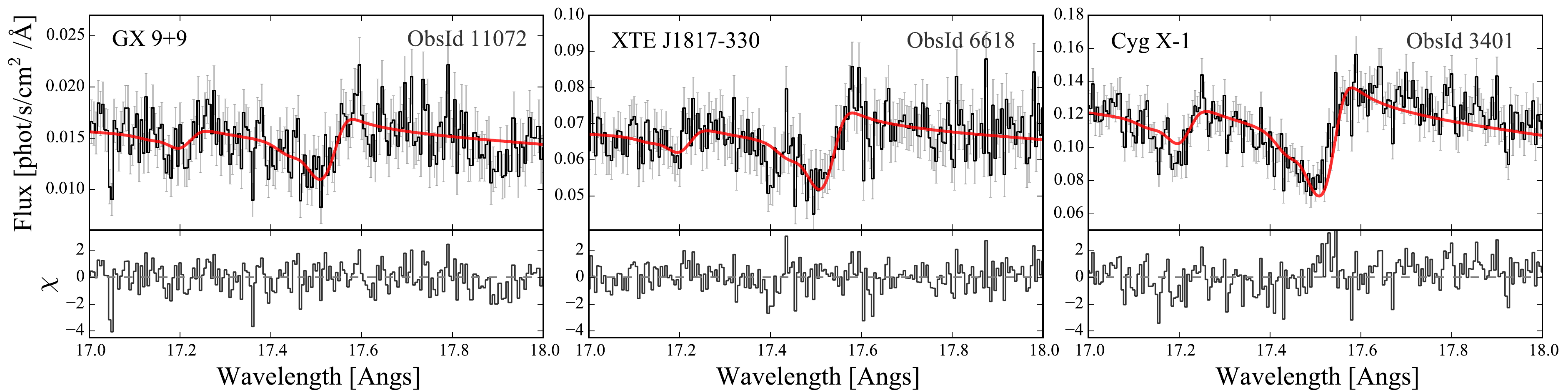}
	\caption{
	Fits to the Fe-L edge from three Galactic X-ray binaries in the {\sl Chandra} archive, using the extinction template in Figure~\ref{fig:FeL}.
	}
	\label{fig:FeL_fits}
\end{center}
\end{figure*}

For now, we would like to illustrate the ability of the Fe-L edge model shown in Figure~\ref{fig:FeL} to fit current observations of X-ray binaries in the {\sl Chandra} archive.  Figure~\ref{fig:FeL_fits} shows a local power law fit in the 16 - 19~\AA\ range, with the Fe-L extinction template in Figure~\ref{fig:FeL}, for GX 9+9, XTE J1817-330, and Cyg X-1.  Table~\ref{tab:FeLfits} gives the equivalent dust mass, $\NH$ column under the fiducial dust abundance assumptions, and the $\NH$ column from \citet[][hereafter GG15]{Gatuzz2015}.  
The columns in GG15 are measured by fitting broad band spectra (11 - 25\AA) with their absorption model \texttt{ISMabs}.  This model is similar to \tbnew, with the difference that it incorporates revised atomic data from not only the neutrals, but also from the single and double ionized species.  However, for Fe-L only, they implement the experimental cross section from solid metallic iron \citep{Kortright2000}.

%We note that, in the case of XTE J1817-330, Fe appears relatively over-abundant, causing the equivalent $\NH$ column (inferred from the Fe edge alone) to disagree with the GG15 value (inferred from a broad band fit).  
Note that the equivalent $\NH$ columns, inferred from the Fe edge alone, can disagree with the GG15 value, which is inferred from a broad band fit.  However, the relative amounts of ferric dust between the three objects in Table~\ref{tab:FeLfits} do agree with the Fe abundance trends seen in \citet{Gatuzz2015}.  In particular for the case of XTE J1817-330, Fe appears relatively over-abundant for the $\NH$ value by a factor of 10.  
There also appears to be slight misalignment between the Fe-L template and the Cyg X-1 data.  This could indicate a difference in dust grain composition, as the relative strengths and positions of the Fe-L$_2$ and -L$_3$ extinction bumps can depend on, for instance, the relative amount of Fe$^{3+}$ in the mineral \citep{vanAken2002}.  \citet{Hanke2009} also noticed a $540 \pm 230$~km~s$^{-1}$ blue shift between the Cyg X-1 Fe-L edge and the pure solid Fe cross section of \citet{Kortright2000}.  We reserve comparisons between Fe-L absorption templates for various materials for a future work.  However, better data is needed to distinguish between different grain types, because overall the quality of the fits is pretty much the same.  Additional grain parameters that affect absorption edge fine structure, such as shape \citep{HD2015} and crystalline structure \citep{Lee2010}, would require greater energy resolution than is currently available.

\begin{table}
\centering
\caption{Fe-L edge fits to three X-ray binaries}
\label{tab:FeLfits}
\begin{tabular}{l c c c}
	\hline
					& {Dust mass} 			& {Equiv $\NH$}		& {ISMabs $\NH$}	\\
	{\bf Object}		& {\tiny ($10^{-4}$~g~cm$^{-2}$)}	& {\tiny ($10^{22}$~cm$^{-2}$)}	& {\tiny ($10^{22}$~cm$^{-2}$)} \\
	\hline
	GX 9+9			& $0.34 \pm 0.04$			& $0.38 \pm 0.04$ 		& $0.315 \pm 0.021$	\\
	XTE J1817-330		& $0.27 \pm 0.03$			& $0.30 \pm 0.03$		& $0.038 \pm 0.020$	\\
	Cyg X-1			& $0.52 \pm 0.03$			& $0.58 \pm 0.03$		& $0.623 \pm 0.014$	\\
	\hline
\end{tabular}
\end{table}

%%- - - - - - - - - - - - - - - - - - - - - - - - - - - - - - - - - - - - - - - - - - - - - - - - - - - %%
\section{Relevance to dust obscured objects}
\label{sec:Discussion}

There are several rules of thumb that one can apply in order to gauge the relative significance of dust absorption and scattering in the X-ray regime.  First, the total optical depth to absorption should be linearly proportional to dust mass, unless the dust grains are large enough to shield inner portions of the grain from X-ray light.  As long as most grains have a radius $a \lsim 0.3~\mum$ the total absorption cross section should not change significantly.  
Second, the ISM optical depth to scattering depends roughly on $\tau_{\rm sca} \propto M_d \rho_g a$, where $\rho_g$ is the density of the dust grain material.  This is because $\tau_{\rm sca}^{\rm RGD} = N_d \sigma_{\rm sca}$ with $N_d \propto M_d / \rho_g a^3$ and $\sigma_{\rm sca} \propto a^4 \rho_g^2$, in the RGD regime.  In the soft X-ray, where the Mie scattering cross section tends to flatten out, one can expect the scattering optical depth to fall significantly with grain size, $\tau_{\rm sca}^{\rm Mie} \propto M_d/\rho_g a^3$.

Finally, the need to include dust scattering as part of the ISM extinction model depends 
%strongly on two factors: the imaging resolution of the X-ray telescope and the location of the dust along the sight line.  The previous text has focused primarily on the effect of telescope imaging resolution.  We turn now to the latter concern.
most strongly on the location of the dust along the sight line.  Interstellar dust always lies at intermediate positions in the sight line so as to produce a dust scattering halo that will be resolved by sub-arcminute imaging telescopes.  However, dust that is closer to the X-ray source produces a more compact scattering halo.  When taking this to the logical extreme of dust that is intrinsic to the X-ray source, the scattering component will not be resolved by any available X-ray telescope.  In that case, there will be no scattering contribution to the extinction.  This fact is highly relevant for dust obscured active galactic nuclei (AGN), young stars, and winds from evolved stellar companions of X-ray binaries.  Models for the cold material intrinsic to these objects should incorporate {\sl absorption only}.

There is one more caveat to this fact -- it assumes isotropic distribution of the dust around the point source.  Scattering from non-symmetric dusty material intrinsic to the source will alter the extinction properties if the angular size of the dust structure is smaller than the typical X-ray scattering angle.  For a spherical dust clump with radius $R$, the distance between the X-ray source and the clump $d$, and the 2$\sigma$ width of the Gaussian approximation to the differential cross section, $\charsig \sim 10' (a/0.1~\mum)^{-2} (E/{\rm keV})^{-1}$, the clumps must be smaller than 
\begin{equation}
\label{eq:R}
	R \lsim 6~{\rm pc}\ (d/{\rm kpc})\ (a / 0.1~\mum)^{-1}\ (E/{\rm keV})^{-1}	 
\end{equation}
for scattering to contribute to extinction.  For the small angles involved, the respective covering factor for a spherical dust cloud is roughly 
\begin{equation}
\label{eq:fcov}
	1 - f_{\rm cov} \approx \exp \left[ -\frac{(R/d)^2}{4 \charsig^2} \right]
\end{equation}
so that $\tau_{\rm ext} = \tau_{\rm abs} + (1-f_{\rm cov}) \tau_{\rm sca}$.  
The dusty torus of obscured AGN or dust disks around T-Tauri stars might produce this effect.  In addition, AGN often exhibit changes in the amount of X-ray obscuration with time, which has been interpreted as variable absorption from clumpy material \citep[e.g.,][]{Risaliti2002}.  
%Dust clumps in AGN narrow-line regions would have to be $\lsim 0.6$~pc in radius to alter absorption edge structure as well.
%
%A search for variability in the low-energy end of the absorption edge fine structure could give clues to the location and size of the absorbing structures.
A recent survey of AGN variability with RXTE detected eclipse events within eight objects \citep{Markowitz2014}. The clouds in their study have an average diameter of 0.25~light days and a distance from the supermassive black hole $\sim 10$ - 100 of light days, which implies $f_{\rm cov} \approx 0.16$ to 1.  Thus, high resolution spectroscopy of some AGN might also exhibit variability in the low-energy (scattering) side of the absorption edge fine structure if the clumps contain dust.

Finally, quasar absorption line systems (ALSs) provide a unique case.  Some of these systems, mainly Mg II absorbers, are thought to have ISM columns $\NH \gsim 10^{21}$~cm$^{-2}$, similar to damped Lyman-$\alpha$ systems (DLAs).  They have also been shown to be dusty \citep[e.g.,][]{York2006,Menard2009}.  
% York2006, but string size is too long for mnras bib style
The Mg II and DLA systems seen in absorption are likely $< 10$~kpc in size, and thus would not be resolved with dust scattering \citep{Corrales2012}.  However, the respective covering fraction for the quasars whose light they intercept is small enough to cause extinction.  Taking cosmological expansion into account, for ALSs with a physical radius $R_a$ at redshift $\zals$, the covering fraction will be 
\begin{equation}
\label{eq:fcos}
	1 - f_{\rm cov} \approx \exp \left[ - \frac{R_a^2 (1+\zals)^2}{4 \charsig^2 {\chi_d}^{2}} \right]
\end{equation}
where 
\begin{equation}
\label{eq:chi}
	\chi_d = \int_{\zals}^{z_{\rm q}} \frac{c\ dz}{H(z)}
\end{equation}
for X-ray sources at $z_q$ and flat cosmologies with expansion $H(z)$.  For most situations, 
%\begin{align}
%\label{eq:Iobs}
%	I_{\nu,0}^{\rm scat} (\alpha) = \ & 
%	F_{\nu,0}^{\rm src} \ \times \\
%	&
%	\int_0^{z_s} n_c (z) \ 
%	\frac{(1+z)^2 }{ x^2 } \ 
%	\frac{d\sigma_{\nu,z}}{d\Omega} \left( \frac{\alpha} {x} \right) \ 
%	\frac{ c \ dz }{ H(z) } \nonumber
%\end{align}
\begin{align}
\label{eq:fcos_values}
	f_{\rm cov} \sim \ & 10^{-3} (1 + \zals)^2\ 
		(a/0.1~\mum)^{2}\ (E/{\rm keV})^{2} \\\
		&
		 \times \ \left( \frac{R_a}{10~{\rm kpc}} \right)^{2}\ 
		\left( \frac{\chi_d}{{\rm Gpc}} \right)^{-2} \nonumber
\end{align}
meaning that absorption edge structure from ALSs will likely be altered by scattering.

%%- - - - - - - - - - - - - - - - - - - - - - - - - - - - - - - - - - - - - - - - - - - - - - - - - - - %%
\section{Summary and conclusions}
\label{sec:Conclusions}

X-ray extinction by dust scattering is a necessary physical component for properly fitting low and high resolution X-ray spectra.  We find that for high resolution telescopes such as \Chandra, the $\NH$ value measured from X-ray absorption will be overestimated by a minimum of 25\%, regardless of true ISM column.  For the most obscured sight lines, $\NH > 10^{23}$~cm$^{-2}$, the photon index for the X-ray spectral energy distribution and the calculation for unattenuated flux can be underestimated by 10-50\%.  The overall effects change depending on the imaging resolution of the telescope, however.  For \XMM\  and \Swift, which have imaging resolution on the order of $10''$, the amount of scattered light recaptured within in a point source aperture will depend strongly on the location of the dust.  The built in XSPEC module \texttt{dust} does not properly account for this, and we advise users to obtain the XSPEC model from Smith et al. (2016, in prep)\footnote{https://github.com/AtomDB/xscat}.  For low-resolution instruments like \nustar\ and Astro-H, the effects of dust scattering will be much more modest.  

Dust scattering also alters the appearance of absorption edge fine structure, most dramatically for the C-K, O-K, and Fe-L edges.  Including the scattering component boosts the total optical depth to extinction and therefore must be included to accurately measure the elemental abundances, even if the edge shape does not change significantly.  When fitting absorption edges, it is best to normalize by dust mass instead of $\NH$, which requires many assumptions about ISM metal abundances and depletion factors.

Finally, extinction by dust scattering can be ignored for objects whose X-ray attenuation comes mainly from cold gas intrinsic to the source.  This includes extinction from dust intrinsic to AGN and stellar disks, regardless of the dust-to-gas ratios in those environments.  However, asymmetric gas coverage or scattering from sufficiently small dust clumps will cause dust scattering to contribute to total extinction, modified by some covering factor.  Thus one might expect photoelectric absorption edges to change shape for AGN that exhibit variable absorption, if the variations come from dust clumps.  In addition, quasar absorption line systems are also small enough to contribute to unresolved scattering extinction.

%%- - - - - - - - - - - - - - - - - - - - - - - - - - - - - - - - - - - - - - - - - - - - - - - - - - - %%

\vspace{0.25in}
We would like to thank Norbert Schulz, Mike Nowak, and Randall Smith for many useful discussions.  The codes used to model dust extinction from the IR to X-ray are open-source and publicly available at github.com/eblur/dust \citep{dustrepo}.  Modules for fitting the high resolution edge structures in this paper, using either an XSPEC local model or a custom ISIS fit function, are available at github.com/eblur/ismdust.

\bibliography{references}

\end{document}